\documentclass[10pt,leqno]{amsart}

\pdfoutput=1

\usepackage{amsmath,amssymb,amsthm}
\usepackage{graphicx}
\usepackage{xcolor}
\usepackage{paralist}
\usepackage{csquotes}
\usepackage{hyperref}
\usepackage{fancyhdr}
\usepackage{etoolbox}

\usepackage{tikz}
\usetikzlibrary{arrows.meta,positioning,shapes.geometric,shapes.multipart,fit,backgrounds}

\usepackage{adjustbox}
\usepackage{multirow}
\usepackage{float}
\usepackage{subcaption}

\usepackage{pgfplots}
\pgfplotsset{compat=1.18}
\begin{document}

\title{Fine-Tuning Small Language Models for Solution-Oriented
Windows Event Log Analysis} 


\author{Siraaj Akhtar}
\thanks{Corresponding author: Siraaj.Akhtar@hud.ac.uk} 

\author{Saad Khan}

\author{Simon Parkinson}

\address{School of Computing and Engineering, University of Huddersfield, Huddersfield, United Kingdom}
\email{Siraaj.Akhtar@hud.ac.uk, saad.khan@hud.ac.uk, S.Parkinson@hud.ac.uk}

\maketitle

\maketitle
\maketitle


\begin{abstract}
Large language models (LLMs) have shown promise for event log analysis, but their high computational requirements, reliance on cloud infrastructure, and security concerns limit practical deployment. In addition, most existing approaches focus only on the identification of the problem and do not provide actionable remediation. Small language models (SLMs) present a light-weight alternative that can be fine-tuned for a specific purpose and hosted locally. This paper investigates whether SLMs, when fine-tuned for a specific task, can serve as a practical alternative for event log analysis while also generating solutions. We first create a large-scale synthetic Windows event log dataset that contains remediation actions using a high-performing LLM. We then fine-tune multiple SLMs and LLMs using the LoRA parameter-efficient fine-tuning technique and evaluate their performance by comparing with expert assessment. The results show that the dataset accurately reflects real-world scenarios and that fine-tuned SLMs consistently outperform LLMs in identifying issues and providing relevant remediation, while requiring fewer computational resources.
\end{abstract}

\bigskip

\section{Introduction}
In previous research for event log analysis, researchers have demonstrated that fine-tuning is one of the most effective methods to adapt language models to specific tasks~\cite{mandakath2024root, chen2024high, pedroso2025anomaly, lim2025adapting}. However, fine-tuning comes with two main limitations, which are high resource requirements and the significant amount of time required to train language models. In addition, LLMs also have significant requirements. To meet these requirements, organisations purchase GPUs or, in most cases, they rent them; both incur high costs. Furthermore, the tools are restricted to only identifying problems and do not introduce new features. This makes it difficult to justify their use for event log analysis tasks, as most non-LLM-based tools can accurately complete the same tasks, are free, and commonly accessible desktop machines meet their resource requirements. Using cloud GPUs to operate the tools is not a suitable option, and if the fine-tuning dataset contains an oranisation's event logs, using them to fine-tune it is also not suitable. This is because the logs will be stored on the cloud provider’s servers~\cite{khan2025managing}, which will be a cyber security threat and a breach of data confidentiality. Additionally, the motivation for fine-tuning open-source LLMs is that closed-source LLMs (such as GPT and Claude) are unsuitable because the logs are sent to the service providers, but if the fine-tuned models are deployed in cloud-based environments, then they will also share the original limitation.

SLMs (Small Language Models), as implied by their name, are smaller language models than LLMs, and thus can be deployed on regular machines. Researchers have demonstrated that they can achieve results comparable to those of their LLM counterparts in other research fields~\cite{javaheripi2023phi, hu2024minicpm, ma2025adaptivelog}. In our domain, only one article has been written that uses SLMs for event log analysis, but this is in combination with LLMs, not independently~\cite{yan2024collaborate}. Consequently, our primary objective is to fill this research gap and assess whether SLMs can perform as well as, or even better than, LLMs when fine-tuned for event log analysis with a focus on providing solutions to problems identified, as this was also not explored in prior research.

To achieve this objective, we generate a dataset that represents real-world data, contains different types of events, includes logs which represent system health, is in the required format for fine-tuning and includes solutions to logs that were identified as problematic. We seek to achieve this by providing the model with real-world logs that cover a range of activities, before generation. Alongside this, we have secondary objectives within fine-tuning itself, which are to reduce resource requirements and time taken, and aim to achieve this using the LoRA technique, whilst still producing an accurate tool. 

We conduct two experiments. In the first experiment, we provide the LLM (Claude 3.7 Sonnet) with a structured prompt and example logs, and ask it to generate 10,000 logs in the correct format while providing solutions to problems that were identified, as the literature mentions that having more than 5000 dataset entries is the best practice to ensure successful fine-tuning~\cite{vieira2024much}. In the second experiment, we used this dataset to fine-tune the selected models and then provided them with a testing dataset to evaluate their outputs. To evaluate both experiments, we asked nine experts (from academia and industry) to complete a questionnaire to assess whether the generated dataset is suitable and evaluate the performance of the language models. 

The novel contributions of this study are:


\begin{itemize} 
\item \textbf{A solution-aware synthetic Windows event-log dataset for fine-tuning.} We construct a large-scale fine-tuning dataset by using real-world Windows log samples to prompt a high-performing LLM to generate realistic logs and paired problem-identification and remediation outputs, addressing the common limitation that prior datasets/tools focus on detection without providing an actionable solution. 
\item \textbf{Fine-tuning to include solutions in event-log analysis.} We fine-tune several SLMs and LLMs for Windows event-log analysis and build on prior research by not only identifying problems but also providing remediation solutions. 
\item \textbf{Human-in-the-loop benchmarking of SLMs vs LLMs on correlated log groups.} We evaluate fine-tuned models on a real-world Windows testing set with correlated log groups (via DepCOMM) and use expert assessment to judge correctness, sufficiency of remediation steps, and practical interpretability enabling an evidence-based comparison of SLM/LLM trade-offs specifically for event-log diagnostics. 
\end{itemize}

In addition to the results, we demonstrate that providing the LLM with examples allowed it to produce realistic event logs, which include a range of different types of activities, and the experts affirmed that the solutions provided contain sufficient information regarding how to resolve problems that were identified. Furthermore, we also found that using the LoRA fine-tuning technique significantly reduced the time taken for fine-tuning, allowing us to fine-tune on a lower specification GPU, and the outputs of the models were still of high quality. In addition, the experts also found that the SLMs were able to rival, and at times, outperform the LLMs regarding identifying problems and providing correct and detailed solutions. This was evident when the SLM versions of Bloom and Gemma significantly outperformed the LLM versions. 

This paper is structured as follows: Firstly, we review literature related to this topic, which highlights the importance of our study, in Section~\ref{litrev}. Following this, we outline our research methodology and motivation in Section~\ref{methodology}. We then transition into our two experiments, first generating our dataset in Section~\ref{datagen} and fine-tuning the LLMs and SLMs in~\ref{finetuning}. Finally, we present our results and discuss their implications, limitations and future work in Section~\ref{discussion}.

\section{Related work}
\label{litrev}
\subsection{Fine-tuning}
Previous research demonstrates that fine-tuning significantly improves results for various tasks, including event log analysis. Due to this, we begin by analysing articles which highlight this trend.  

In the research article,~\cite{mandakath2024root}, the authors compare performance for root cause analysis between fine-tuned models (MistralLite and Mixtral-87B) and GPT-4. They successfully demonstrate that MistralLite was able to outperform GPT-4 and even the larger model, Mixtral-87b. As a result of this, we can conclude that fine-tuning is an essential tool to address the limitations of open-source models and that smaller models are better suited for fine-tuning because fewer parameters need to be changed. Another study,~\cite{chen2024high}, also justifies the use of fine-tuning. The authors aim to improve efficiency for deep learning parsers with fine-tuning and online parsing. As was the case with the previous article, the results demonstrated an improvement over the prior techniques. A limitation of this study is that the parsing is online, which presents security concerns, as the logs will be vulnerable to malicious actors who could intercept them, which can be addressed in future research. 

In a recent study,~\cite{pedroso2025anomaly}, the authors aim to adapt their tool for cloud-based environments. To achieve this, Llama-3 8B is fine-tuned, and they achieve near-optimal scores. To further improve upon this study, future work can explore methods to reduce the significant costs and time required for fine-tuning. In addition to using additional techniques, like event log correlation, to further improve accuracy. 

Addressing this limitation,~\cite{ren2024clogllm}, uses prompt engineering in combination with fine-tuning. It is important to note that prior research focused on these techniques individually, and this study expands upon them by integrating both techniques. As was the case with all the previous articles we have analysed, the study demonstrated that their tool was able to significantly improve in performance and outperform a model from the GPT series. Despite this, the limitation of high costs, both in time and computation resource, remains. The study,~\cite{lim2025adapting}, aims to evaluate solutions to this salient limitation by using PEFT (parameter-efficient fine-tuning) techniques. This technique differs from traditional fine-tuning as it modifies specific parameters of the model rather than all of them, resulting in training taking place at a faster rate and a reduction in the resources required. However, the datasets and models generated do not represent various issues because the study is restricted to anomaly detection. Future work can address this by evaluating PEFT techniques for a comprehensive range of event log analysis tasks by fine-tuning with a comprehensive dataset. In addition, they can expand to not just identifying binary anomalies but also identifying the cause of the issues and providing solutions to them.

\subsection{Small language models (SLMs)}
Deploying LLMs locally requires powerful hardware; therefore, most organisations tend to host them in the cloud. This poses security risks and violations of policies and regulations. Recently, researchers have explored SLMs as an alternative to address the resource limitation. The research study,~\cite{javaheripi2023phi}, evaluates the Phi series of SLMs and argues that they can, at times, perform better than larger models. They achieve this by using high-quality data similar to that found in `` textbooks, and their results justify their claim. Similarly, the research article,~\cite{hu2024minicpm}, aims to produce an SLM which can rival its LLM counterparts. They use two techniques (model wind tunnel experiments and the WSD Learning Rate Scheduler) to efficiently and optimally modify how data is sent to the model and its parameters, respectively, during training. Their results show that they achieved their objective, highlighting the importance of additional techniques to bridge the gap in performance. 

Moving on, the authors, in the study,~\cite{wang2024comprehensive}, conduct a survey on the literature of SLMs and find that they are especially important in areas where security is a priority. Hosting LLMs requires a significant number of resources, and as a result, most developers decide to deploy their models using cloud infrastructure, which poses security concerns as the data is vulnerable to cyber-attacks and violates certain policies and laws which prevent data from being shared with external parties. They highlight how SLMs have not been explored in a wide range of domains, which they suggest as an avenue for future research. 

Finally, we look at a study that is within our domain,~\cite{ma2025adaptivelog}. In this article, the authors combine the use of SLMs with LLMs; the LLM is only used for responses where the SLM is likely not to understand the logs. The results showed that the required resources were reduced while the results remained highly accurate. However, it is essential to note that this solution still requires a significant amount of resources, as an LLM is still used. Future work can utilise methods such as providing high-quality datasets (as mentioned in the article based on the Phi series of models), which would allow SLMs to be used instead of larger models.

\subsection{Event log correlation}
Event log correlation refers to collecting all logs that relate to an event. Some logs may be considered benign if analysed individually, but if they are present with other logs, they could clearly indicate that malicious activities have taken place, or are taking place. As a result of this, it is important to integrate event log correlation into LLM-based tools, as previously, this has been overlooked. 

To begin, in the study,~\cite{farshchi2018metric}, the authors aim to produce a tool which can successfully detect software errors using event log-based correlation. They collated logs within a certain time period that possessed single or multiple "cloud platform metrics"; their results demonstrated that the tool was able to accurately identify faults. A similar approach is explored in the research article,~\cite{dubuc2021real}. The authors use Apache Spark and Elastic Search to correlate events based on a range of attributes, and they conclude that Elastic Search is a superior technique because it is quicker and more resource-efficient. However, a limitation of these two approaches is that some events may have logs which take place at different time periods and therefore would go undetected by this tool.

We now move on to studies which could address this limitation. In the study,~\cite{khan2017causal}, the authors develop a tool which correlates event logs and can be used by ordinary users. They use ARM to establish repeated events. The logs are then verified to be connected if there are matches between the objects within the logs. Finally, the logs are passed through an algorithm, and a cause-and-effect percentage is generated. The results show that the tool has a low processing time and consistently has high accuracy scores. A comparable approach is provided in the article,~\cite{li2022feature}, where the authors use ‘'Kendall rank correlation and Granger causality tests'’ to correlate logs which are repeated and remove them. The results show that they successfully manage to improve results, and future work can expand to include cyber attacks, as this study is restricted to redundant logs.

A different approach is taken in the study,~\cite{kobayashi2017mining}, where a graph-based approach is used to eliminate logs. They only use the logs which are in direct proximity and remove the ones that are spatially nearby. Their results demonstrate the practicality of their system on real-world data, but, like the previous study, it is limited to redundant logs. A comprehensive solution is provided in the research article,~\cite{wang2022end}. Firstly, alerts are correlated based on various attributes. Afterwards, the alerts are correlated, plotted in a graph, and Monte Carlo Tree Search is used to identify missing logs for that activity, based on previous alert patterns. The results show improvement compared to prior research. A limitation of this study is that novel events may not be detected, as the tool is restricted to previous patterns. 

Finally, in the study,~\cite{xu2022depcomm}, the authors used a graph-based approach to group logs. Their approach is an improvement over prior research, as they merge similar nodes and edges into ``communities''. A community is a group of logs which relate to a single activity, and the authors highlight an improvement in results when compared to prior research.

\subsection{Solutions for limited context window}
Applying event correlation to LLM-based correlation could lead to hallucinations and even errors due to the large amount of information that exceeds the limited context windows of the models. As a result of this, we aim to explore some solutions from other domains to inspire our solution to this problem.

To begin, we examine studies which suggest repeating instructions could resolve this issue. In the articles ~\cite{yu2023self, yousuf2025can}, the authors introduce this important technique and repeat instructions when the model is responding to a large prompt, as the models are likely to hallucinate if they are not reminded of the task that they should be performing. Future research can utilise this technique to ensure that the model is consistently reminded of the instructions, thereby reducing the likelihood of memory drift.

Moving on to solutions that reduce the amount of information sent to the prompt at a time, in the research article,~\cite{hosseini2024efficient}, the authors demonstrate that sending summaries of the text or even just the first and last sentences of it can significantly improve results. Comparably, in the study,~\cite{ho2025arcmemo}, the authors propose externally storing data. The authors demonstrate that their model produced better results, even when compared to top-performing models, such as Claude 3.7 Sonnet. Future research could use this methodology to analyse large pieces of text by sending data in chunks. 

Finally, we look at retrieval-based solutions. In the article,~\cite{xu2023retrieval}, the authors aim to investigate whether searching for external information can serve as an alternative to large context windows and fine-tuning. The authors demonstrate that an LLM with a 75\% less context window was able to produce high-quality outputs when compared to a larger, fine-tuned model. A comparative study is conducted in the article,~\cite{zeng2024structural}, where the authors aim to investigate different solutions to the limitation of the limited memory capacity of LLMs. They find that the best-performing solution differs between the tasks. However, iterative retrieval consistently performs the best. Nonetheless, applying this technique to our domain would be challenging. Log format changes over time, making it difficult to obtain, update and store logs for retrieval.

\subsection{Research gap}
The first subsection highlighted the importance of fine-tuning within our domain, but also the limitations of costs, resource requirements, and that the solutions are restricted to identifying problems without any additional functionality. We then highlighted the effectiveness of SLMs in other domains, the importance of event log correlation, and potential solutions to sending large amounts of information to LLMs in other domains. This demonstrates the importance of our study to extend the functionality of LLM-based event log analysis tools (by providing mitigation steps, rather than only problem identification) and assessing whether SLMs are suitable, which addresses the cost and resource limitations.

\section{Methodology}

\newlength{\boxwidth}
\setlength{\boxwidth}{\dimexpr\columnwidth-12pt\relax}

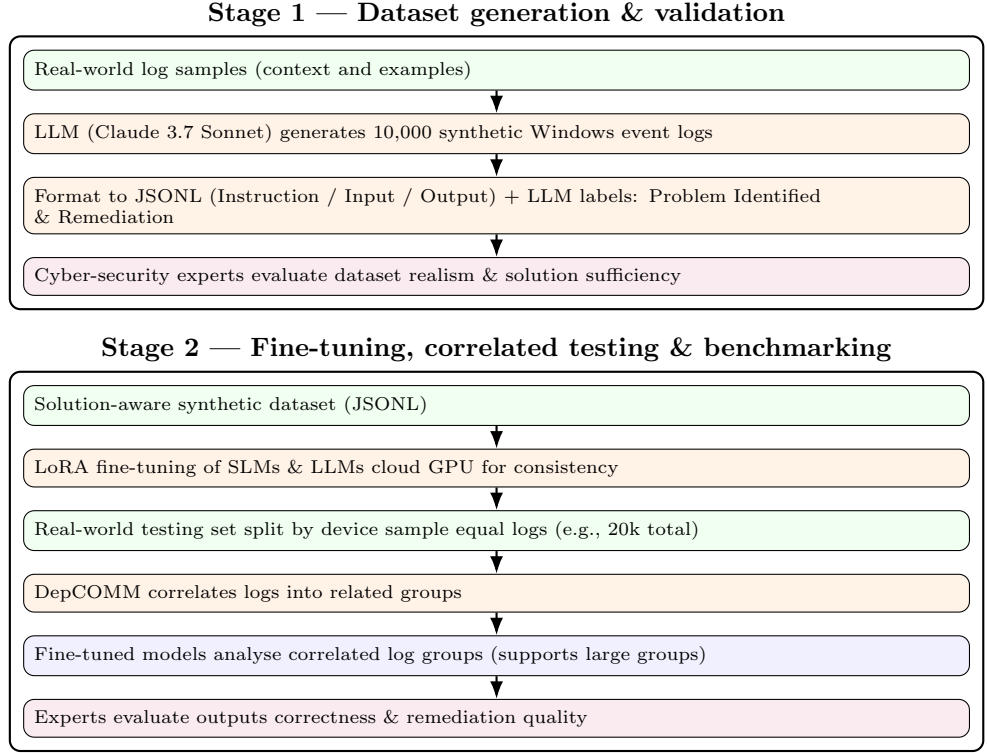
\begin{figure}[t]

\begin{tikzpicture}[
  font=\scriptsize,
  node distance=4mm,
  >=Latex,
  box/.style={
    draw,
    rounded corners,
    align=left,
    inner sep=4pt,
    text width=\boxwidth
  },
  proc/.style={box, fill=blue!6},
  data/.style={box, fill=green!6},
  tool/.style={box, fill=orange!10},
  human/.style={box, fill=purple!8},
  arrow/.style={->, thick},
  stage/.style={draw, rounded corners, thick, inner sep=5pt}
]

\node[stage, label={[font=\bfseries]above:Stage 1 — Dataset generation \& validation}] (S1) {

\begin{tikzpicture}[node distance=3mm]

\node[data] (realexamples)
{Real‑world log samples (context and examples)};

\node[tool, below=of realexamples] (llmgen)
{LLM (Claude 3.7 Sonnet) generates 10,000 synthetic Windows event logs};

\node[tool, below=of llmgen] (formatlabel)
{Format to JSONL (Instruction / Input / Output) + LLM labels: Problem Identified \\ \& Remediation};

\node[human, below=of formatlabel] (experts1)
{Cyber‑security experts evaluate dataset realism \& solution sufficiency};

\draw[arrow] (realexamples) -- (llmgen);
\draw[arrow] (llmgen) -- (formatlabel);
\draw[arrow] (formatlabel) -- (experts1);

\end{tikzpicture}
};

\node[stage, below=8mm of S1,
      label={[font=\bfseries]above:Stage 2 — Fine‑tuning, correlated testing \& benchmarking}] (S2) {

\begin{tikzpicture}[node distance=3mm]

\node[data] (ftdata)
{Solution‑aware synthetic dataset (JSONL)};

\node[tool, below=of ftdata] (loraft)
{LoRA fine‑tuning of SLMs \& LLMs cloud GPU for consistency};

\node[data, below=of loraft] (testprep)
{Real‑world testing set split by device sample equal logs (e.g., 20k total)};

\node[tool, below=of testprep] (correlate)
{DepCOMM correlates logs into related groups};

\node[proc, below=of correlate] (analyze)
{Fine‑tuned models analyse correlated log groups (supports large groups)};

\node[human, below=of analyze] (experts2)
{Experts evaluate outputs correctness \& remediation quality};

\draw[arrow] (ftdata) -- (loraft);
\draw[arrow] (loraft) -- (testprep);
\draw[arrow] (testprep) -- (correlate);
\draw[arrow] (correlate) -- (analyze);
\draw[arrow] (analyze) -- (experts2);

\end{tikzpicture}
};

\end{tikzpicture}

\caption{Methodology overview: (Stage 1) solution‑aware synthetic dataset generation with expert validation; (Stage 2) LoRA fine‑tuning of SLMs/LLMs and expert‑evaluated testing on correlated real‑world log groups.}
\label{fig:method_overview}
\end{figure}

\label{methodology}
\subsection{Overview} 
As illustrated in Figure~\ref{fig:method_overview}, this study has two main stages. The first stage is to create a dataset for fine-tuning. Most datasets are small, do not cover a wide range of events and attacks, do not include logs that incorporate system health issues, and do not provide solutions to identified problems. Recent studies have demonstrated that LLMs can be used to produce suitable datasets~\cite{zhang2024logfilm, li2025anomalygen}. In the second stage, we use this dataset to fine-tune a range of SLMs and LLMs and compare their performance to determine if the SLMs are an alternative for our domain. Figure~\ref{fig:method_overview} illustrates an overview of this paper.




\subsection{Motivation for study and objectives}
In our survey article~\cite{akhtar2025llm}, we frequently mentioned that closed-source LLMs can not be used as they send data to the service provider, which is a breach of security. However, even small LLMs have significant computational requirements and cannot be deployed within most organisations because their machines do not adhere to the resource requirements. Due to this, cloud providers are used, which is also unsuitable, as the sensitive event log data will be sent to the cloud providers.

As a result of this, our primary objective is to ensure that our solution is secure. Recent research~\cite{javaheripi2023phi, hu2024minicpm, wang2024comprehensive} has demonstrated that SLMs (which can be deployed on most local machines) could be used as an alternative to LLMs when trained for a specific task. In this study, we focus on Windows event logs. We train SLMs and LLMs and then compare their results for the task, and intend to highlight that the SLMs can achieve results that are equal to or similar to their LLM counterparts.

LLMs are known for generating human-like text, but event log researchers have not exploited this. Prior tools are restricted to identifying issues and not providing solutions to them. It is important to fill this research gap because non-LLM-based tools can accurately identify issues; if our tools only provide the same functionality, then LLM-based event log analysis could be considered redundant. 

\subsection{Experiment environment}
As we mentioned in the previous subsection, LLMs have significant resource requirements and cannot be deployed locally. Due to this, we decided to conduct our main experiment (fine-tuning and testing it) in a cloud-based environment to ensure consistency. We utilised Runpod~\cite{runpod2025}, which allows users to rent a GPU and fine-tune directly. We rented a GPU as it allowed us to use a specific technique of fine-tuning. The GPU used was the A100 PCIe, as it fulfils the resource requirements of all the models selected. This does not contradict our objectives, as the SLMs used could be deployed locally, but if we decided to use the cloud for LLMs and local machines for the SLMs, our study would be considered methodologically inconsistent. 

\subsection{Correlation Tool}
The objective of our experiment is not to correlate logs; however, most research regarding the use of LLMs to perform various event log analysis tasks analyses logs individually. Some logs by themselves may not appear malicious, but if they are accompanied by other logs, then they could clearly represent attacks. A simple example of this is a DDoS attack, where an action is repeated with the intent to cause a service to be overwhelmed and crash. Most LLM-based tools would not detect this attack, as the activity itself may not be malicious, even though the larger group clearly indicates that a DDOS attack is taking place. The tool we selected for correlation is DepCOMM~\cite{xu2022depcomm}, as it is from a highly cited research article and covers a range of event logs (such as Windows, Android and Linux), which means that it is suitable for our future research projects. In addition, it accepts the parsed format of logs, which our fine-tuning and testing datasets are comprised of. 

\subsection{Testing details}  

To perform testing, we selected a real-world dataset containing 20,000 logs (double the size of fine-tuning data) and covers a wide range of activities~\cite{mehulkatara_windows_event_log_2025}. However, the original dataset contains over 600,000 logs. We calculated that if we were to perform analysis on the original dataset using all models, it would take approximately 209 days (6 months and 27 days) and cost \$7085.43. As a result, we split the data into seven smaller datasets, based on the seven machines within the dataset. We calculated that our experiments could be performed within our budget and time window if the dataset was equal to or less than 20,000 logs, so we reduced each dataset to 2857 logs. Following this, DepCOMM correlates the logs, the models perform analysis for the logs within the dataset, and finally, the experts evaluate the models' responses, which are then used to determine if the SLMs were able to identify problems and provide solutions in a manner equal to or better than the LLMs. Figure~\ref{fig:method_overview} illustrates this methodology.

\subsection{Expert evaluation}
Our results cannot be calculated through mathematical equations as there is no ground-truth, and solutions to problems identified could differ, whilst still being correct. Other researchers used experts within the domain of cyber-security to resolve this issue~\cite{hermann2024gpt, ahmed2024prompting, mandakath2024root}. Their objective was to determine if the LLM-based tools would perform analysis in a similar manner to theirs. Our experiment also utilises this evaluation method; we selected nine experts in the domain from industry and academia to evaluate the suitability of our dataset and to assess and compare the responses of the SLMs and LLMs.

\section{Dataset Generation}
\label{datagen}
\subsection{Justification for generating a dataset}
To begin, we analysed a range of event log datasets (including those used in other research based on using LLMs for event log analysis) and identified three main limitations that render them unsuitable for fine-tuning. The first is size, for fine-tuning, a minimum of 1000 entries is mentioned; however, to achieve better results, it is better to have over 5000~\cite{vieira2024much}. Most datasets, such as LogHub, are limited to 2000 logs. In addition to this, the second limitation is that the datasets are not in JSONL format~\cite{nunez2025mldataforge, manibalan2024commandsense, jarin2024using}; the structure should be [Instruction: …, Input: …, Output: …]. Furthermore, prior datasets do not include the steps that can be taken to resolve issues within them. As a result of all these limitations, it was incumbent upon us to develop our own dataset. 

In our previous studies, we frequently mentioned that event logs are sensitive data, and as a result, we were unable to request event logs from our institution and other organisations with which we work alongside to form our dataset. Additionally, if we were to use our logs from our own machines, it would not be sufficient, as it would contain a limited range of activities, and to produce an effective tool, a comprehensive dataset is required. Due to this, we reached the conclusion that generating a dataset would be the most suitable option to obtain a dataset that meets our requirements.

\subsection{Why are we using LLMs to generate our dataset?}
Within the literature, we found research articles which highlight that powerful closed-source LLMs could be used to generate high-quality datasets. These, as we have frequently mentioned, are unsuitable for event log analysis itself, but they are suitable for generating datasets ~\cite{nunez2025mldataforge, manibalan2024commandsense, jarin2024using}, as they will not encounter sensitive data. Furthermore, it would not be possible to manually label the entries with their solutions due to time constraints, and the quality of the solutions could suffer due to fatigue. We decided to use Claude 3.7 Sonnet as it can be accessed through the OpenRouter API, and it was shown in our previous paper~\cite{akhtar2025evaluating} to be the best performing model. 

\subsection{Prompt engineering}
To ensure that the model produces a dataset that meets our requirements, we need to provide it with suitable prompts. Notably, we also found (in our previous study) that prompts need to have the following qualities, which would also apply to this task: they need to be clear, relevant and specific~\cite{wang2024prompt}, specify an identity to the model~\cite{shao2023character} and provide it with context. As our dataset generation contains two stages: generating a dataset and labelling it, then putting it into the correct format, we need two different prompts. The first prompt (for the first stage) is ``You are a Windows event log generator. Generate 9,000 realistic Windows event logs (including performance from performance monitor, system logs, and application events) as JSON objects. Each log entry should always have the following fields: Date/Time, Event ID, Application and Description. All columns should not include any commas. Include logs as separate entries and do not add additional columns.'' The second prompt (for the second stage) is ``You are an event log analyser. Reply only with the text to fill in the output field. Reply only in the format: Problem Identified..., How to resolve: …,''.



Moving on to our implementation details, which are also represented in ~\ref{fig:method_overview}. We use Claude 3.7 Sonnet (through the OpenRouter API) to generate 10,000 synthetic event logs in JSON format, using the prompt which was mentioned in the previous subsection. Secondly, we used DepCOMM to correlate the Windows LogHub dataset, as this step ensures that correlated groups of logs are represented in the dataset. This is important because we want our tool to have the ability to analyse both individual logs and those in groups. The third step is a script which puts the log into the correct format. It automatically fills in the Instruction column with our instruction for fine-tuning, as this needs to be the same for all logs, then it places the log (this will occur for each log) in the Input column. The final part of this step is where Claude will generate the output statement, again using the relevant prompt mentioned previously. In the final step, cyber-security experts will review the dataset and determine if it meets their requirements, which will be elaborated on in Sections~\ref{discussion} of this paper.

\subsection{Dataset characteristics}
Prior research has the limitation of generating data which does not accurately represent real-world scenarios. As we mentioned, context is important~\cite{shao2023character}, so in addition to the prompt itself, we also send various example logs from real-world systems, which include system health, various attacks and normal activity. The primary and distinguishing characteristic of this dataset is that it includes the solutions to the problematic logs and log groups. Furthermore, we include logs that are relevant to a range of event log analysis tasks, as previous datasets are restricted to one task. Prior research and datasets did not include this, but it is an important feature to prevent redundancy in LLM-based event log literature and harness the full capabilities of the models. 

\subsection{Cost and time}
The total costs from using the OpenRouter interface to request Claude 3.7 Sonnet for both steps were \$23.02, and the total time taken was 14 hours.

\subsection{Example pipeline}





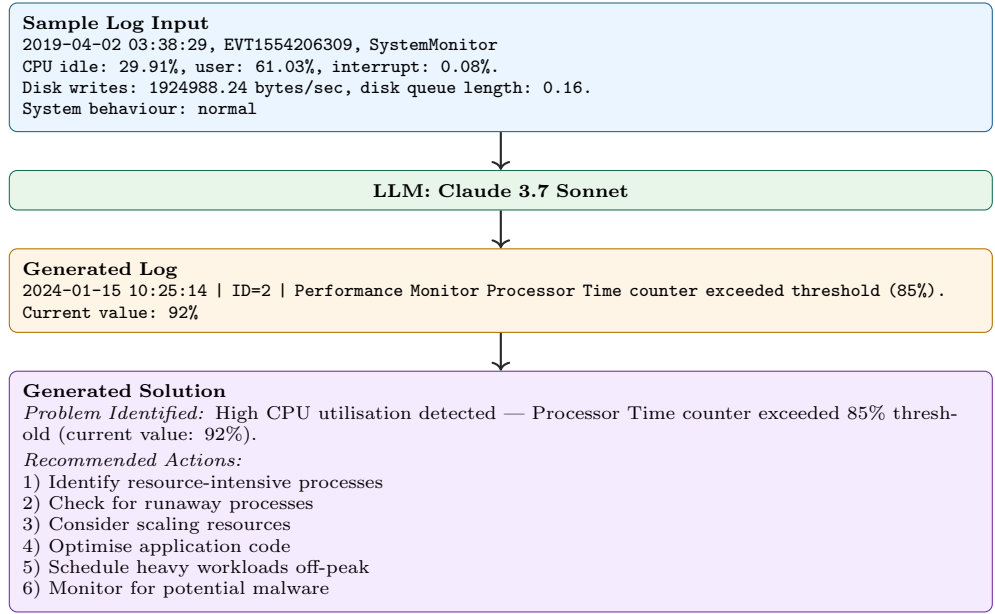
\begin{figure}[t]

\definecolor{boxA}{RGB}{33,97,140}
\definecolor{boxAfill}{RGB}{235,245,255}

\definecolor{boxB}{RGB}{24,111,83}
\definecolor{boxBfill}{RGB}{232,245,233}

\definecolor{boxC}{RGB}{183,110,0}
\definecolor{boxCfill}{RGB}{255,245,230}

\definecolor{boxD}{RGB}{112,48,160}
\definecolor{boxDfill}{RGB}{245,235,255}

\definecolor{arrowcol}{RGB}{40,40,40}

\begin{tikzpicture}[
    node distance=0.5cm,
    box/.style={
        draw,
        rectangle,
        rounded corners=3pt,
        align=left,
        inner sep=5pt,
        text width=\columnwidth,
        font=\scriptsize
    },
    centrebox/.style={
        draw,
        rectangle,
        rounded corners=3pt,
        align=center,
        inner sep=5pt,
        text width=\columnwidth,
        font=\scriptsize
    },
    arrow/.style={
        ->,
        thick,
        draw=arrowcol
    }
]

\node[box, draw=boxA, fill=boxAfill] (A) {
\textbf{Sample Log Input}\\
\texttt{2019-04-02 03:38:29, EVT1554206309, SystemMonitor}\\
\texttt{CPU idle: 29.91\%, user: 61.03\%, interrupt: 0.08\%.}\\
\texttt{Disk writes: 1924988.24 bytes/sec, disk queue length: 0.16.}\\
\texttt{System behaviour: normal}
};

\node[centrebox, draw=boxB, fill=boxBfill, below=of A] (B) {
\textbf{LLM: Claude 3.7 Sonnet}
};

\node[box, draw=boxC, fill=boxCfill, below=of B] (C) {
\textbf{Generated Log}\\
\texttt{2024-01-15 10:25:14 | ID=2 | Performance Monitor Processor Time counter exceeded threshold (85\%). Current value: 92\%}
};

\node[box, draw=boxD, fill=boxDfill, below=of C] (D) {
\textbf{Generated Solution}\\
\textit{Problem Identified:} High CPU utilisation detected — Processor Time counter exceeded 85\% threshold (current value: 92\%).\\[2pt]
\textit{Recommended Actions:}\\
1) Identify resource‑intensive processes\\
2) Check for runaway processes\\
3) Consider scaling resources\\
4) Optimise application code\\
5) Schedule heavy workloads off‑peak\\
6) Monitor for potential malware
};

\draw[arrow] (A) -- (B);
\draw[arrow] (B) -- (C);
\draw[arrow] (C) -- (D);

\end{tikzpicture}

\caption{Dataset generation pipeline using an LLM for log synthesis and remediation guidance.}
\label{fig:datasetpipeline}
\end{figure}

In Figure \ref{fig:datasetpipeline}, we use example data to demonstrate how we generated our dataset. To begin, a sample real-world log is sent to the LLM (Claude 3.7 Sonnet). In our example, there is a log relating to system health, which shows that the system is in a healthy state because CPU utilisation is at an acceptable level. The objective of this is for Claude to generate logs which represent both good and poor system health. Following this, we show that Claude was successfully able to generate a dataset entry that represents poor system health, as the CPU utilisation is above the threshold. Subsequently, Claude was able to provide a detailed six-step plan of how to resolve the issue, which will teach the models how to respond to such issues if they were to occur. The amount of detail provided in the solution is important to ensure that when the dataset is used for fine-tuning, the model can understand it and provide correct solutions. If the amount of detail is minimal, the models may not understand how to resolve the issue and produce incorrect solutions to the problems that were identified. 

\section{Fine-tuning}
\label{finetuning}
\subsection{Motivation}
In our previous research articles, we highlighted that fine-tuning is the best-performing LLM technique. Other techniques can help to improve results, but if they are used without fine-tuning, results significantly decrease. As a result of this, we selected it as our main tool for this experiment. Fine-tuning has significant resource requirements, and as we mentioned previously, cloud platforms are typically used, which are costly. In addition, the time taken to perform fine-tuning is high~\cite{church2021emerging}. Due to this, we will also attempt to reduce costs, computational requirements, and time taken while maintaining the accurate identification of problems and their solutions.

Furthermore, we aim to evaluate if SLMs can be used as an alternative to LLMs, as we also mentioned in our methodology section. We intend to select a range of SLMs and LLMs, compare them and highlight that SLMs can perform equal to or even better than LLMs when they are trained to identify issues and provide solutions for a specific event log analysis task, in this case, Windows event logs. This will pave the way for future research where multiple SLM iterations can be developed, for different types of event logs, and then they can be linked together and analysed by one tool. This is how we aim to achieve the second novel contribution of this paper.

\subsection{Model selection}
To begin, we will relay the SLMs that were chosen for this study, as well as justify our choices. The first model is BTLM 3b; this model was chosen as research demonstrated that it was able to perform like a 7b (the size of small LLMs) model~\cite{dey2023btlm}. The second model that was chosen was Bloom 4b. This is primarily because it also has a 7b version, which will allow us to compare SLM and LLM models from the same series directly. Likewise, Gemma 3 4b was selected as it is the SLM form of the first LLM selected. 

Moving on, as we mentioned, Gemma 3 7b is the first LLM selected, as in our previous research article~\cite{akhtar2025llm}, the results illustrated that it was the best-performing LLM. Bloom 7b was selected as it has an SLM form and was shown in prior research to perform well~\cite{duong2024bloomllm}. Finally, Mistral 7b was selected as it was also a high-performing model from our previous comparison paper. All the LLMs selected are 7b as in our previous papers; these models were shown to outperform those that were larger than them, and larger models would require more powerful GPUs, which would not allow us to be able to complete our experiments within our budget. 

\subsection{LoRA fine-tuning}
To achieve our aim of reducing resource requirements and time taken for fine-tuning, we selected the LoRA fine-tuning technique. This is important, as prior event log analysis research has yet to evaluate this technique for multiple event log analysis tasks (it is only restricted to anomaly detection), despite it being shown to be effective in other domains. LoRA fine-tuning differs from regular fine-tuning as it modifies low-rank matrices rather than the full matrices, which would be significantly larger. Researchers have shown that this is more effective than regular fine-tuning, as its performance does not decrease and the model is less likely to forget information~\cite{shuttleworth2024lora, wang2023lora, tian2024hydralora}.

\subsection{Implementation details}
Now, we will discuss our implementation details. We will begin with the batch size, which was 16 for the SLMs and 4 for the LLMs, except Bloom 7b, as it was reduced to 2, as it caused the GPU to crash. All models had 3 epochs, and they differed in maximum token length as each model had specific requirements. The rest of the implementation was the same for all models.

\subsection{Testing details}
The testing details, however, differ as there is a pre-detection method for the SLMs to aid them in achieving better results. This feature is designed to detect a range of attacks, including ransomware, brute-force, privilege escalation, and various software failures. This is to enhance their results and requires an insignificant amount of processing power, so it will not conflict with our objectives. Other than this, the testing script used is the same (for SLMs and LLMs) and sends a system prompt requesting the models to reply with whether a problem was identified; if one is, then they are asked to provide a solution. 

In addition, we introduced a method to handle large groups of logs, as they are currently unable to do so due to their limited context window. This idea was inspired by the following research articles~\cite{zhang2022summn, zhou2025llm}, which use a similar method but for large amounts of text. This method is executed if there are more than 7 logs in a group, where it will send the previous response of the model and ask it to update its response if there is any additional information; if not, it will keep it the same, and only the final response will be emitted for that group. As was the case with our dataset, experts will be used to evaluate the responses of the models, and this analysis will be used to compare the models, leading to a conclusion on whether we reached our objectives.

\section{Results and discussion}

\label{discussion}
In this section, we will discuss our results. All of the completed forms, as well as the complete summary charts (which contain all of the results), are publicly \href{https://drive.google.com/drive/folders/1Xbp5DDlFcBSEThpnRAf7_SC2ZI4g_93h?usp=sharing}{available}.

\subsection{Results summary tables}
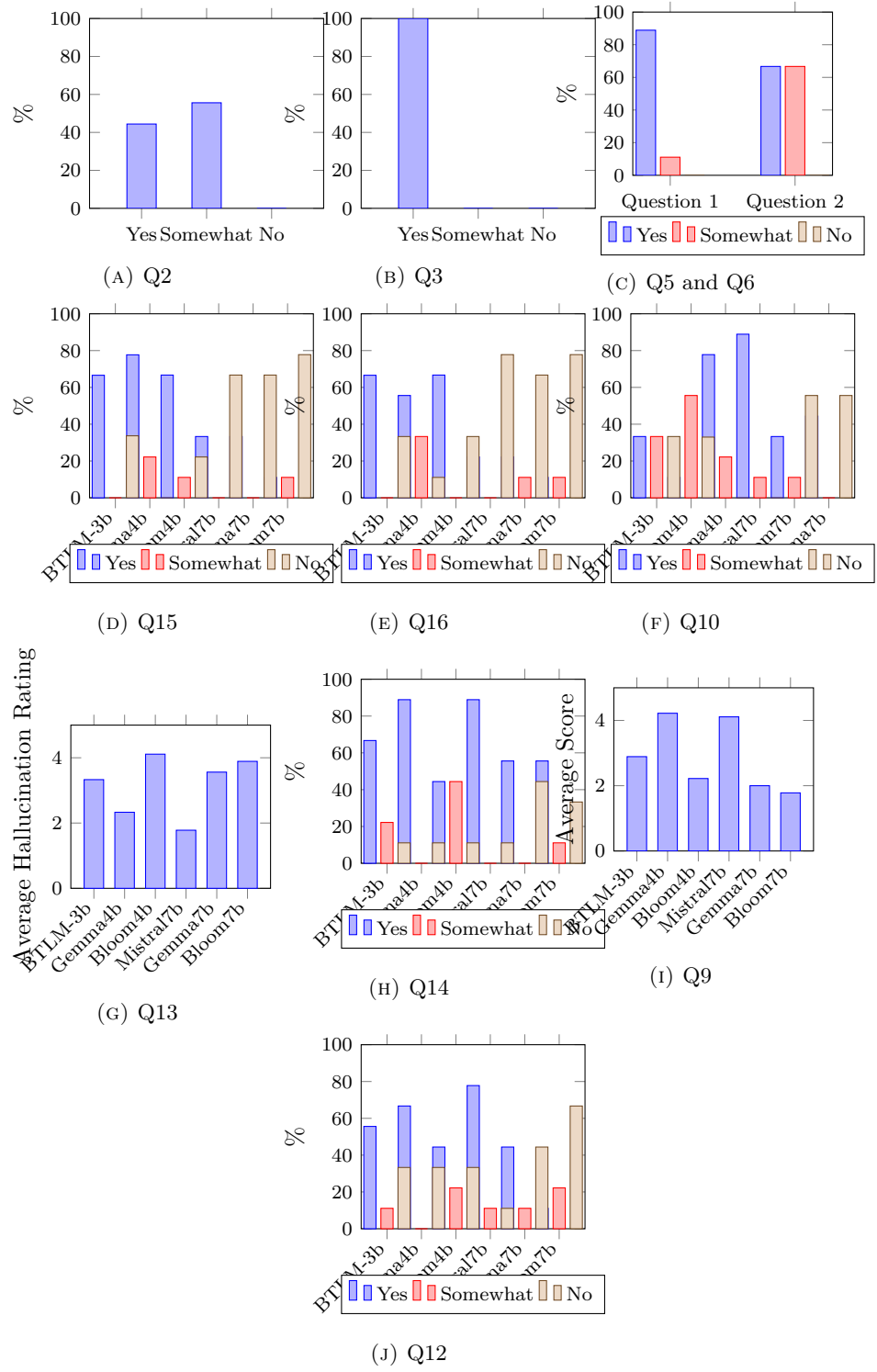
\begin{figure*}[!ht]
\begin{subfigure}[!ht]{0.3\textwidth}
\begin{tikzpicture}
\begin{axis}[
    ybar,
    bar width=12pt,
    height=5cm,
    ymin=0,
    ymax=100,
    ylabel={\%},
    symbolic x coords={Yes, Somewhat, No},
    xtick=data,
    xticklabel style={font=\footnotesize},
    yticklabel style={font=\footnotesize},
    enlarge x limits=0.4,
    scale=0.80,transform shape
]

\addplot coordinates {
    (Yes, 44.4)
    (Somewhat, 55.6)
    (No, 0)
};

\end{axis}
\end{tikzpicture}
\caption{Q2}
\label{fig:chart1}
\end{subfigure}
\begin{subfigure}[!ht]{0.3\textwidth}
\begin{tikzpicture}
\begin{axis}[
    ybar,
    bar width=12pt,
    height=5cm,
    ymin=0,
    ymax=100,
    ylabel={\%},
    symbolic x coords={Yes, Somewhat, No},
    xtick=data,
    xticklabel style={font=\footnotesize},
    yticklabel style={font=\footnotesize},
    enlarge x limits=0.4,
    scale=0.80,transform shape
]

\addplot coordinates {
    (Yes, 100)
    (Somewhat, 0)
    (No, 0)
};

\end{axis}
\end{tikzpicture}
\caption{Q3}
\label{fig:chart2}
\end{subfigure}
\begin{subfigure}[!ht]{0.3\textwidth}
\begin{tikzpicture}
\begin{axis}[
    ybar,
    bar width=8pt,
    height=5.5cm,
    ymin=0,
    ymax=100,
    ylabel={\%},
    symbolic x coords={Question 1, Question 2},
    xtick=data,
    xticklabel style={font=\footnotesize},
    yticklabel style={font=\footnotesize},
    legend style={
        font=\footnotesize,
        at={(0.5,-0.25)},
        anchor=north,
        legend columns=3
    },
    enlarge x limits=0.3,
    scale=0.60,transform shape
]

\addplot coordinates {
    (Question 1, 88.9)
    (Question 2, 66.7)
};

\addplot coordinates {
    (Question 1, 11.1)
    (Question 2, 66.7)
};

\addplot coordinates {
    (Question 1, 0)
    (Question 2, 0)
};

\legend{Yes, Somewhat, No}

\end{axis}
\end{tikzpicture}
\caption{Q5 and Q6}
\label{fig:chart3}
\end{subfigure}
\begin{subfigure}[!ht]{0.3\textwidth}
\begin{tikzpicture}
\begin{axis}[
    ybar,
    height=6cm,
    ymin=0,
    ymax=100,
    ylabel={\%},
    symbolic x coords={
        BTLM-3b,
        Gemma4b,
        Bloom4b,
        Mistral7b,
        Gemma7b,
        Bloom7b
    },
    xtick=data,
    xticklabel style={rotate=45, anchor=east, font=\footnotesize},
    yticklabel style={font=\footnotesize},
    legend style={
        font=\footnotesize,
        at={(0.5,-0.25)},
        anchor=north,
        legend columns=3
    },
    bar width=5pt,
    enlarge x limits=0.15,
     scale=0.60,transform shape
]

\addplot coordinates {
    (BTLM-3b,66.6)
    (Gemma4b,77.7)
    (Bloom4b,66.7)
    (Mistral7b,33.3)
    (Gemma7b,33.3)
    (Bloom7b,11.1)
};

\addplot coordinates {
    (BTLM-3b,0)
    (Gemma4b,22.2)
    (Bloom4b,11.1)
    (Mistral7b,0)
    (Gemma7b,0)
    (Bloom7b,11.1)
};

\addplot coordinates {
    (BTLM-3b,33.7)
    (Gemma4b,0)
    (Bloom4b,22.2)
    (Mistral7b,66.7)
    (Gemma7b,66.7)
    (Bloom7b,77.8)
};

\legend{Yes, Somewhat, No}

\end{axis}
\end{tikzpicture}
\caption{Q15}
\label{fig:chart4}
\end{subfigure}
\begin{subfigure}[!ht]{0.3\textwidth}
\begin{tikzpicture}
\begin{axis}[
    ybar,
    height=6cm,
    ymin=0,
    ymax=100,
    ylabel={\%},
    symbolic x coords={
        BTLM-3b,
        Gemma4b,
        Bloom4b,
        Mistral7b,
        Gemma7b,
        Bloom7b
    },
    xtick=data,
    xticklabel style={rotate=45, anchor=east, font=\footnotesize},
    yticklabel style={font=\footnotesize},
    legend style={
        font=\footnotesize,
        at={(0.5,-0.25)},
        anchor=north,
        legend columns=3
    },
    bar width=5pt,
    enlarge x limits=0.15,
     scale=0.60,transform shape
]

\addplot coordinates {
    (BTLM-3b,66.6)
    (Gemma4b,55.6)
    (Bloom4b,66.7)
    (Mistral7b,22.2)
    (Gemma7b,22.2)
    (Bloom7b,11.1)
};

\addplot coordinates {
    (BTLM-3b,0)
    (Gemma4b,33.3)
    (Bloom4b,0)
    (Mistral7b,0)
    (Gemma7b,11.1)
    (Bloom7b,11.1)
};

\addplot coordinates {
    (BTLM-3b,33.3)
    (Gemma4b,11.1)
    (Bloom4b,33.3)
    (Mistral7b,77.8)
    (Gemma7b,66.7)
    (Bloom7b,77.8)
};

\legend{Yes, Somewhat, No}

\end{axis}
\end{tikzpicture}
\caption{Q16}
\label{fig:chart5}
\end{subfigure}
\begin{subfigure}[!ht]{0.3\textwidth}
\begin{tikzpicture}
\begin{axis}[
    ybar,
    bar width=5pt,
    height=6cm,
    ymin=0,
    ymax=100,
    ylabel={\%},
    ylabel style={font=\small},
    symbolic x coords={
        BTLM-3b,
        Bloom4b,
        Gemma4b,
        Mistral7b,
        Bloom7b,
        Gemma7b
    },
    xtick=data,
    xticklabel style={rotate=45, anchor=east, font=\footnotesize},
    yticklabel style={font=\footnotesize},
    legend style={
        font=\footnotesize,
        at={(0.5,-0.25)},
        anchor=north,
        legend columns=3
    },
    enlarge x limits=0.15,
     scale=0.60,transform shape
]

\addplot coordinates {
    (BTLM-3b,33.3)
    (Bloom4b,11.1)
    (Gemma4b,77.8)
    (Mistral7b,88.9)
    (Bloom7b,33.3)
    (Gemma7b,44.4)
};

\addplot coordinates {
    (BTLM-3b,33.3)
    (Bloom4b,55.6)
    (Gemma4b,22.2)
    (Mistral7b,11.1)
    (Bloom7b,11.1)
    (Gemma7b,0)
};

\addplot coordinates {
    (BTLM-3b,33.3)
    (Bloom4b,33)
    (Gemma4b,0)
    (Mistral7b,0)
    (Bloom7b,55.6)
    (Gemma7b,55.6)
};

\legend{Yes, Somewhat, No}

\end{axis}
\end{tikzpicture}
\caption{Q10}
\label{fig:chart6}
\end{subfigure}
\begin{subfigure}[!ht]{0.3\textwidth}
\begin{tikzpicture}
\begin{axis}[
    ybar,
    height=5.5cm,
    ymin=0,
    ymax=5,
    ylabel={Average Hallucination Rating},
    symbolic x coords={
        BTLM-3b,
        Gemma4b,
        Bloom4b,
        Mistral7b,
        Gemma7b,
        Bloom7b
    },
    xtick=data,
    xticklabel style={rotate=45, anchor=east, font=\footnotesize},
    yticklabel style={font=\footnotesize},
    bar width=8pt,
    enlarge x limits=0.15,
     scale=0.60,transform shape
]

\addplot coordinates {
    (BTLM-3b,3.33)
    (Gemma4b,2.33)
    (Bloom4b,4.11)
    (Mistral7b,1.78)
    (Gemma7b,3.56)
    (Bloom7b,3.89)
};

\end{axis}
\end{tikzpicture}
\caption{Q13}
\label{fig:chart7}
\end{subfigure}
\begin{subfigure}[!ht]{0.3\textwidth}
\begin{tikzpicture}
\begin{axis}[
    ybar,
    height=6cm,
    ymin=0,
    ymax=100,
    ylabel={\%},
    symbolic x coords={
        BTLM-3b,
        Gemma4b,
        Bloom4b,
        Mistral7b,
        Gemma7b,
        Bloom7b
    },
    xtick=data,
    xticklabel style={rotate=45, anchor=east, font=\footnotesize},
    yticklabel style={font=\footnotesize},
    legend style={
        font=\footnotesize,
        at={(0.5,-0.25)},
        anchor=north,
        legend columns=3
    },
    bar width=5pt,
    enlarge x limits=0.15,
     scale=0.60,transform shape
]

\addplot coordinates {
    (BTLM-3b,66.7)
    (Gemma4b,88.9)
    (Bloom4b,44.4)
    (Mistral7b,88.9)
    (Gemma7b,55.6)
    (Bloom7b,55.6)
};

\addplot coordinates {
    (BTLM-3b,22.2)
    (Gemma4b,0)
    (Bloom4b,44.4)
    (Mistral7b,0)
    (Gemma7b,0)
    (Bloom7b,11.1)
};

\addplot coordinates {
    (BTLM-3b,11.1)
    (Gemma4b,11.1)
    (Bloom4b,11.1)
    (Mistral7b,11.1)
    (Gemma7b,44.4)
    (Bloom7b,33.3)
};

\legend{Yes, Somewhat, No}

\end{axis}
\end{tikzpicture}
\caption{Q14}
\label{fig:chart8}
\end{subfigure}
\begin{subfigure}[!ht]{0.3\textwidth}
\begin{tikzpicture}
\begin{axis}[
    ybar,
    height=5.5cm,
    ymin=0,
    ymax=5,
    ylabel={Average Score},
    symbolic x coords={
        BTLM-3b,
        Gemma4b,
        Bloom4b,
        Mistral7b,
        Gemma7b,
        Bloom7b
    },
    xtick=data,
    xticklabel style={rotate=45, anchor=east, font=\footnotesize},
    yticklabel style={font=\footnotesize},
    bar width=8pt,
    enlarge x limits=0.15,
     scale=0.60,transform shape
]

\addplot coordinates {
    (BTLM-3b,2.89)
    (Gemma4b,4.22)
    (Bloom4b,2.22)
    (Mistral7b,4.11)
    (Gemma7b,2.00)
    (Bloom7b,1.78)
};

\end{axis}
\end{tikzpicture}
\caption{Q9}
\label{fig:chart9}
\end{subfigure}
\begin{subfigure}[!ht]{0.3\textwidth}
\begin{tikzpicture}
\begin{axis}[
    ybar,
    height=6cm,
    ymin=0,
    ymax=100,
    ylabel={\%},
    symbolic x coords={
        BTLM-3b,
        Gemma4b,
        Bloom4b,
        Mistral7b,
        Gemma7b,
        Bloom7b
    },
    xtick=data,
    xticklabel style={rotate=45, anchor=east, font=\footnotesize},
    yticklabel style={font=\footnotesize},
    legend style={
        font=\footnotesize,
        at={(0.5,-0.25)},
        anchor=north,
        legend columns=3
    },
    bar width=5pt,
    enlarge x limits=0.15,
     scale=0.60,transform shape
]

\addplot coordinates {
    (BTLM-3b,55.6)
    (Gemma4b,66.7)
    (Bloom4b,44.4)
    (Mistral7b,77.8)
    (Gemma7b,44.4)
    (Bloom7b,11.1)
};

\addplot coordinates {
    (BTLM-3b,11.1)
    (Gemma4b,0)
    (Bloom4b,22.2)
    (Mistral7b,11.1)
    (Gemma7b,11.1)
    (Bloom7b,22.2)
};

\addplot coordinates {
    (BTLM-3b,33.3)
    (Gemma4b,33.3)
    (Bloom4b,33.3)
    (Mistral7b,11.1)
    (Gemma7b,44.4)
    (Bloom7b,66.7)
};

\legend{Yes, Somewhat, No}

\end{axis}
\end{tikzpicture}
\caption{Q12}
\label{fig:chart10}
\end{subfigure}
\caption{Compilation of all charts relating to questions and their answers from the expert evaluation.}
\label{fig:allcharts}
\end{figure*}

A large number of questions were asked to the experts, and there were two evaluation methods: the first was a score out of 5, and the second was whether they agreed (yes) or disagreed (no) with the question asked. To aid the reader, we present our results in tables organised by experiment and evaluation method. Table \ref{tab:dataset_question_table} contains all of the responses to the questions relating to our generated dataset to determine whether it is suitable. The results for the fine-tuned models after analysing the testing dataset are split into two tables. Table \ref{tab:finetuning_5star} illustrates where the response to questions is a score out of 5, and Table \ref{tab:fine-tuning_3qs} is where the responses indicate the extent to which the experts agree with the question. In addition to this, Figure \ref{fig:allcharts} compiles all of the figures relating to individual questions, which will be expanded upon in the following sections.

\begin{table}[H]
\small
\begin{tabular}{|c|c|c|c|}
\hline
\textbf{Question} & \textbf{Yes (\%)} & \textbf{Somewhat (\%)} & \textbf{No (\%)} \\
\hline
Q1 & 100 & 0 & 0 \\
Q2 & 44.4 & 55.6 & 0 \\
Q3 & 100 & 0 & 0 \\
Q4 & 100 & 0 & 0 \\
Q5 & 89.9 & 11.1 & 0 \\
Q6 & 66.7 & 33.3 & 0 \\
Q7 & 88.9 & 11.1 & 0 \\
Q8 & 100 & 0 & 0 \\
\hline
\end{tabular}
\caption{Expert evaluation responses for our generated dataset.}
\label{tab:dataset_question_table}
\end{table}

\begin{table}[H]
\small
\resizebox{\columnwidth}{!}{
\begin{tabular}{|c|ccc|ccc|ccc|ccc|ccc|ccc|}
\hline
 & \multicolumn{3}{c|}{\textbf{BTLM-3b}} 
 & \multicolumn{3}{c|}{\textbf{Gemma 4b}} 
 & \multicolumn{3}{c|}{\textbf{Bloom 4b}} 
 & \multicolumn{3}{c|}{\textbf{Mistral 7b}} 
 & \multicolumn{3}{c|}{\textbf{Gemma 7b}} 
 & \multicolumn{3}{c|}{\textbf{Bloom 7b}} \\
\textbf{Q} 
& Yes(\%) & Some(\%) & No(\%)
& Yes(\%) & Some(\%) & No(\%)
& Yes(\%) & Some(\%) & No(\%)
& Yes(\%) & Some(\%) & No(\%)
& Yes(\%) & Some(\%) & No(\%)
& Yes(\%) & Some(\%) & No(\%) \\
\hline

Q10
& 33.3 & 33.3 & 33.3 
& 77.8 & 22.2 & 0 
& 11.1 & 55.6 & 33.3 
& 88.9 & 11.1 & 0 
& 44.4 & 0 & 55.6 
& 33.3 & 11.1 & 55.6 \\

Q11 
& 44.4 & 33.3 & 22.2 
& 77.8 & 22.2 & 0 
& 22.2 & 55.6 & 22.2 
& 77.8 & 22.2 & 0 
& 44.4 & 11.1 & 44.4 
& 22.2 & 11.1 & 66.7 \\

Q12
& 55.6 & 11.1 & 33.3 
& 66.7 & 33.3 & 0 
& 44.4 & 22.2 & 33.3 
& 77.8 & 11.1 & 11.1 
& 44.4 & 11.1 & 44.4 
& 11.1 & 22.2 & 66.7 \\

Q14
& 66.7 & 22.2 & 11.1 
& 88.9 & 11.1 & 0 
& 44.4 & 44.4 & 11.1 
& 88.9 & 11.1 & 0
& 55.6 & 0 & 44.4 
& 33.3 & 11.1 & 55.6 \\

Q15
& 66.7 & 0 & 33.3 
& 77.8 & 22.2 & 0 
& 66.7 & 11.1 & 22.2 
& 22.2 & 11.1 & 66.7 
& 33.3 & 0 & 66.7 
& 11.1 & 11.1 & 77.8 \\

Q16
& 66.7 & 0 & 33.3 
& 66.7 & 0 & 33.3 
& 55.6 & 33.3 & 11.1 
& 22.2 & 0 & 77.8 
& 22.2 & 11.1 & 66.7
& 11.1 & 11.1 & 77.8 \\

\hline
\end{tabular}
}
\caption{Fine-tuned models' expert evaluation scores for yes/somewhat/no questions.}
\label{tab:fine-tuning_3qs}
\end{table}

\begin{table}[H]
\small
\resizebox{\columnwidth}{!}{
\begin{tabular}{|c|c|c|c|c|c|c|}
\hline
\textbf{Question} & \textbf{BTLM-3b} & \textbf{Gemma 4b} & \textbf{Bloom 4b} & \textbf{Mistral 7b} & \textbf{Gemma 7b} & \textbf{Bloom 7b} \\
\hline
Q9  & 2.89 & 4.22 & 2.22 & 4.11 & 2.00 & 1.78 \\
Q13 & 3.33 & 2.33 & 4.11 & 1.78 & 3.56 & 3.89 \\
\hline
\end{tabular}
}
\caption{Fine-tuned models' expert evaluation scores for 1–5 rating questions.}
\label{tab:finetuning_5star}
\end{table}

\paragraph{Questionnaire items}
\begin{itemize}
\item Q1: Does the template entry below meet the correct JSONL format?
\item Q2: Is the instruction for fine-tuning correct?
\item Q3: Are the event logs throughout the dataset comprehensive (represent various system activities, attacks and system health)?
\item Q4: Are problems correctly identified?
\item Q5: Are the solutions to problems identified, correct?
\item Q6: Do they contain enough information for someone to resolve the issue?
\item Q7: In the final entries (last 4) are correlated groups of logs, that vary in sizes. Is the output (problem identified/not and how to resolve) correct?
\item Q8: For the correlated logs, are all relevant events in the specific groups?
\item Q9: In a general sense, how accurate are the models in identifying issues?
\item Q10: Did the outputs make sense?
\item Q11: Were the solutions for problems identified correct?
\item Q12: Is the information given to resolve issues sufficient?
\item Q13: How many of the responses contained hallucinations? (the lower the score, the lower the amount of hallucinations)
\item Q14: If there were hallucinations, were you still able to understand the responses?
\item Q15: Was the model able to handle large groups of event logs?
\item Q16: For large groups of event logs, did the responses take into account all of the events?
\label{fig:questions}
\end{itemize}

\subsection{Suitability of using expert evaluation}
Both the generated dataset and the testing datasets are large. In addition, our testing dataset lacks ground truth, and there is no suitable mechanism to assess whether the outputs are correct and of good quality. Furthermore, the main author of this paper could not evaluate them himself because the results may be biased. Due to these factors, using experts to evaluate experiments of this nature is the only viable option. The experts evaluated the outputs in a general sense, as they could not be expected to analyse the logs in their entirety, which comes with the limitation that the results do not represent the logs as a whole and are instead a thorough observation. Additionally, the experts could experience fatigue as this is a mundane task; because of this, there is a chance that some of their responses could be incorrect. For example, a clear observation is that Gemma 7b and Bloom 7b hallucinated severely, but a small number of experts said they did not. Future researchers can build upon this by significantly reducing the size of testing datasets whilst ensuring that they contain a range of activities. This could be achieved by randomly selecting logs from different time periods within a dataset. Another approach could be dividing the testing dataset into different parts and assigning experts to them rather than the entire dataset. 

\subsection{Evaluation of instruction for fine-tuning}
\begin{figure}[t]
\begin{tikzpicture}
\begin{axis}[
    ybar,
    bar width=12pt,
    height=5cm,
    ymin=0,
    ymax=100,
    ylabel={\%},
    symbolic x coords={Yes, Somewhat, No},
    xtick=data,
    xticklabel style={font=\footnotesize},
    yticklabel style={font=\footnotesize},
    enlarge x limits=0.4
]

\addplot coordinates {
    (Yes, 44.4)
    (Somewhat, 55.6)
    (No, 0)
};

\end{axis}
\end{tikzpicture}
\caption{Distribution of responses for the validity of the fine-tuning instruction.}
\label{fig:instruction}
\end{figure}
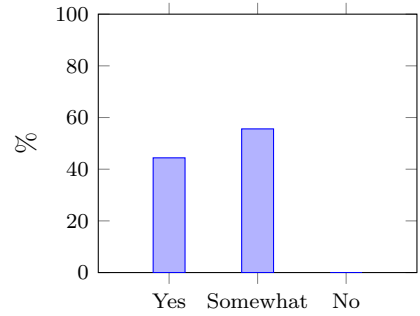

The instruction was consistent across the fine-tuning dataset, as we do not require the fine-tuned models to do any other tasks. The results from the experts were unanimous (highlighted in Figure~\ref{fig:instruction}), which were that the instruction was fit for purpose, as it clearly specified a task and identity to the model, which meets the requirements. However, a considerable number of them mentioned that the use of the term "computational issues” was overly restrictive, as we required the models to identify a broad range of problems, and this statement could restrict it to only system health-related issues. As a result of this, in our future research, we will refine this by either removing the term “computational” or by introducing other terms, such as security and authentication issues, alongside it.

\subsection{Using a LLM to generate a fine-tuning dataset}
Our results clearly demonstrate that using an LLM to generate an event log dataset was successful in our domain. The experts consistently gave the generated dataset high scores; they agreed that it represented a diverse range of activities, correctly identified issues and gave detailed responses on how to resolve issues for problems that were identified. This shows that it meets our primary objective of providing solutions where relevant, and secondly, it accurately represents logs that could be generated with real-world systems. 

\subsection{Evaluation of instruction for fine-tuning}
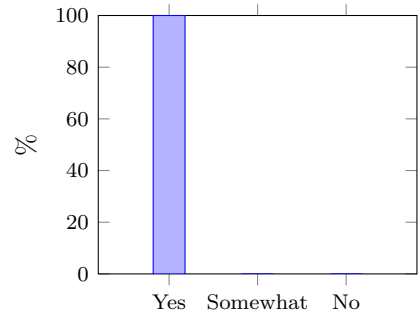
\begin{figure}[t]
\begin{tikzpicture}
\begin{axis}[
    ybar,
    bar width=12pt,
    height=5cm,
    ymin=0,
    ymax=100,
    ylabel={\%},
    symbolic x coords={Yes, Somewhat, No},
    xtick=data,
    xticklabel style={font=\footnotesize},
    yticklabel style={font=\footnotesize},
    enlarge x limits=0.4
]

\addplot coordinates {
    (Yes, 100)
    (Somewhat, 0)
    (No, 0)
};

\end{axis}
\end{tikzpicture}
\caption{Distribution of responses to whether the dataset contained a comprehensive set of logs.}
\label{fig:comprehensive}
\end{figure}

Previously, we mentioned that we provided the LLM with example logs to ensure that it produced realistic logs that covered a range of events. Our results highlighted that this was successful; the main author expanded upon this by further questioning some of the experts. The experts mentioned that the logs were structured consistently, and the type of log attribute appeared at the beginning, and this is illustrated in Figure~\ref{fig:comprehensive}. They were able to look through the logs and see that this attribute regularly changed throughout the dataset. These changes included (but were not limited to) security events, performance monitoring, authentication events, errors, power, service control management and disk management. This is of interest as people are sceptical about using LLMs for generating event log datasets, as they predict that they will not create realistic logs, and they will be limited to a small range of events. Our study shows this concern was not the case. Furthermore, the study highlights the importance of providing the LLM with examples because we were able to ensure that the model produced responses that had the amount of detail that was required, whilst remaining on task and not hallucinating. 

\subsection{Generating problem identified/resolved statements}
\begin{figure}[t]
\begin{tikzpicture}
\begin{axis}[
    ybar,
    bar width=8pt,
    height=5.5cm,
    ymin=0,
    ymax=100,
    ylabel={\%},
    symbolic x coords={Question 1, Question 2},
    xtick=data,
    xticklabel style={font=\footnotesize},
    yticklabel style={font=\footnotesize},
    legend style={
        font=\footnotesize,
        at={(0.5,-0.25)},
        anchor=north,
        legend columns=3
    },
    enlarge x limits=0.3
]

\addplot coordinates {
    (Question 1, 88.9)
    (Question 2, 66.7)
};

\addplot coordinates {
    (Question 1, 11.1)
    (Question 2, 66.7)
};

\addplot coordinates {
    (Question 1, 0)
    (Question 2, 0)
};

\legend{Yes, Somewhat, No}

\end{axis}
\end{tikzpicture}
\caption{Response distribution for two evaluation questions.
Q5: Are the solutions to problems identified, correct?
Q6: Do they contain enough information for someone to resolve the issue?}
\label{fig:problemidentified}
\end{figure}
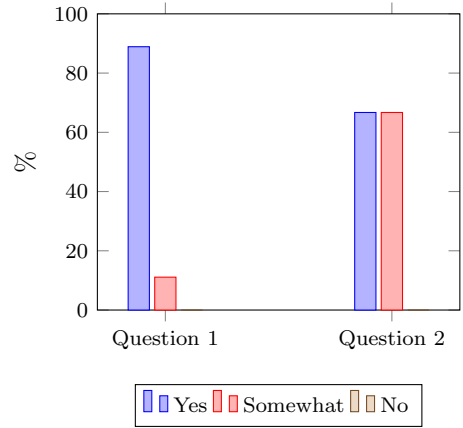
Manually generating output responses for a dataset and identifying problems within the logs would be close to impossible due to the large magnitude of logs and the high risk of fatigue, and as a result, the primary contribution of this paper is to address this concern. The experts agreed that the model was successfully able to identify issues and produce outputs that would help users to resolve them (this is highlighted in Figure~\ref{fig:problemidentified}). This was consistent throughout the dataset (for the different types of event logs), and when the model had to analyse large groups of logs, it was able to do so without a drop in performance. The main author also questioned some of the experts on this topic, and they affirmed that the responses were detailed, did not contain unnecessary information, and the steps given could be understood and utilised by people who lack technical expertise to resolve the issues. In addition, they mentioned the outputs often were a numbered list, which aids users because of high lucidity. The significantly high results within this experiment are due to the speciality of LLMs to generate text-based data, and an argument can be made that this is the event log analysis task that LLMs are best suited to. This is an important conclusion, as it affirms our theory that the large and high-performing LLMs can be used to produce a detailed dataset that includes solutions to the problems within it for event log analysis.

\subsection{Reduction of time taken and resources required with LoRA fine-tuning}
\begin{table}[h]
\begin{tabular}{|l|c|}
\hline
Language Model & Time Taken (h:min) \\ 
\hline
BTLM-3b & 1:13 \\ 
Gemma 4b & 0:53 \\ 
Bloom 4b & 0:35 \\ 
Mistral 7b & 0:35 \\ 
Gemma 7b & 1:13 \\ 
Bloom 7b & 1:01 \\
\hline
\end{tabular}
\caption{Table detailing the time taken for fine-tuning.}
\label{tab:fttimes}
\end{table}

Significant amounts of literature mentioned that fine-tuning for event log analysis tasks took multiple hours~\cite{karlsen2024benchmarking, akhtar2025llm}. However, none of the literature mentions specific time values. This conclusion is also supported by studies in other fields where large numbers of users reported that regular fine-tuning took “between a few hours to seven days”~\cite{huggingface_training_time_fine_tuning, azure_finetuning_duration_qna, glean_fine_tuning_glossary}. As a result of this, one of our primary objectives was to significantly reduce the time taken by using the LoRA fine-tuning technique. Table \ref{tab:fttimes} demonstrates that there was a significant reduction compared to regular fine-tuning for event log analysis-related tasks. Additionally, we successfully executed our experiments on a lower-specification GPU compared to what would be required for regular fine-tuning. This was also even though our dataset entries were larger due to the problem-identified statements and their solutions compared to prior datasets, which were analogous to boolean values. These are important findings, as time can be saved by utilising this technique, as well as costs, as lower-specification GPUs will be less expensive to rent or purchase. Furthermore, the experts found that the responses of the models that could respond correctly were realistic, which highlights that the use of this technique does not reduce accuracy in event log analysis tasks. 

\subsection{Time taken to analyse testing dataset}
\begin{table}[h]
\begin{tabular}{|l|c|}
\hline
Language Model & Time Taken (days:h) \\ 
\hline
BTLM-3b & 3:02 \\ 
Gemma 4b & 0:19 \\ 
Bloom 4b & 3:17 \\ 
Mistral 7b & 0:19 \\ 
Gemma 7b & 0:05 \\ 
Bloom 7b & 1:14 \\
\hline
\end{tabular}
\caption{Table detailing the time taken for testing.}
\label{tab:testtimes}
\end{table}
We will begin by analysing the time taken for the SLMs (rounding up to the nearest hour), which is demonstrated in Table \ref{tab:testtimes}. This was the only category of our results where the LLMs outperformed the SLMs. This is an interesting conclusion, as our study shows the opposite of research papers within other domains, where the SLMs tend to be quicker. This could be explained by the level of complexity of event log analysis. A step to address this in the future could be to train the SLMs for specific types of event logs (even within an operating system), then use an agentic framework to provide a single interface to users.

\subsection{Groups which contain a large number of event logs}
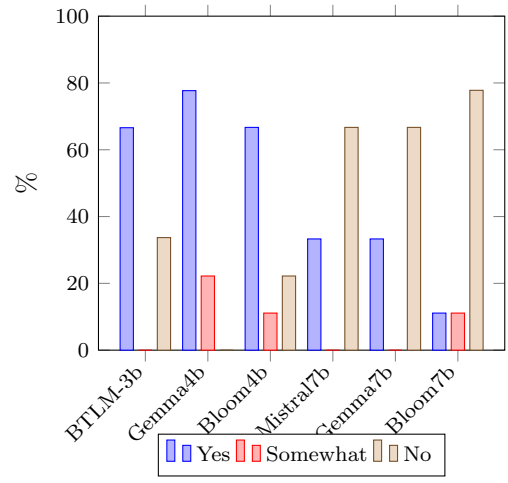
\begin{figure}[t]
\begin{tikzpicture}
\begin{axis}[
    ybar,
    height=6cm,
    ymin=0,
    ymax=100,
    ylabel={\%},
    symbolic x coords={
        BTLM-3b,
        Gemma4b,
        Bloom4b,
        Mistral7b,
        Gemma7b,
        Bloom7b
    },
    xtick=data,
    xticklabel style={rotate=45, anchor=east, font=\footnotesize},
    yticklabel style={font=\footnotesize},
    legend style={
        font=\footnotesize,
        at={(0.5,-0.25)},
        anchor=north,
        legend columns=3
    },
    bar width=5pt,
    enlarge x limits=0.15
]

\addplot coordinates {
    (BTLM-3b,66.6)
    (Gemma4b,77.7)
    (Bloom4b,66.7)
    (Mistral7b,33.3)
    (Gemma7b,33.3)
    (Bloom7b,11.1)
};

\addplot coordinates {
    (BTLM-3b,0)
    (Gemma4b,22.2)
    (Bloom4b,11.1)
    (Mistral7b,0)
    (Gemma7b,0)
    (Bloom7b,11.1)
};

\addplot coordinates {
    (BTLM-3b,33.7)
    (Gemma4b,0)
    (Bloom4b,22.2)
    (Mistral7b,66.7)
    (Gemma7b,66.7)
    (Bloom7b,77.8)
};

\legend{Yes, Somewhat, No}

\end{axis}
\end{tikzpicture}
\caption{Distribution of responses regarding whether the model was able to handle large groups of event logs.}
\label{fig:large_groups}
\end{figure}

\begin{figure}[t]
\begin{tikzpicture}
\begin{axis}[
    ybar,
    height=6cm,
    ymin=0,
    ymax=100,
    ylabel={\%},
    symbolic x coords={
        BTLM-3b,
        Gemma4b,
        Bloom4b,
        Mistral7b,
        Gemma7b,
        Bloom7b
    },
    xtick=data,
    xticklabel style={rotate=45, anchor=east, font=\footnotesize},
    yticklabel style={font=\footnotesize},
    legend style={
        font=\footnotesize,
        at={(0.5,-0.25)},
        anchor=north,
        legend columns=3
    },
    bar width=5pt,
    enlarge x limits=0.15
]

\addplot coordinates {
    (BTLM-3b,66.6)
    (Gemma4b,55.6)
    (Bloom4b,66.7)
    (Mistral7b,22.2)
    (Gemma7b,22.2)
    (Bloom7b,11.1)
};

\addplot coordinates {
    (BTLM-3b,0)
    (Gemma4b,33.3)
    (Bloom4b,0)
    (Mistral7b,0)
    (Gemma7b,11.1)
    (Bloom7b,11.1)
};

\addplot coordinates {
    (BTLM-3b,33.3)
    (Gemma4b,11.1)
    (Bloom4b,33.3)
    (Mistral7b,77.8)
    (Gemma7b,66.7)
    (Bloom7b,77.8)
};

\legend{Yes, Somewhat, No}

\end{axis}
\end{tikzpicture}
\caption{Distribution of responses for whether the responses took into account all of the logs in the log groups.}
\label{fig:all_events}
\end{figure}
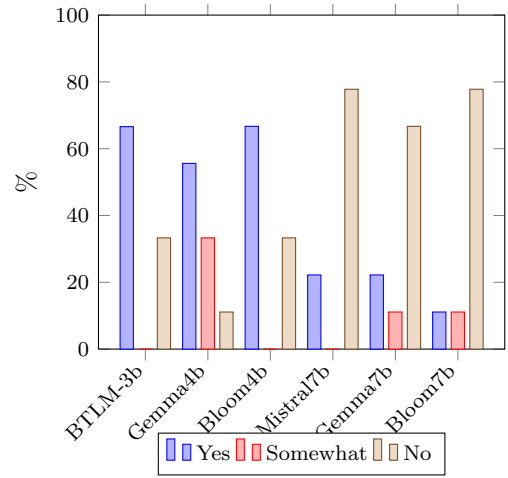

Another important evaluation metric was whether the models could analyse large groups of logs. As we mentioned previously, there are logs that, when analysed individually, will not indicate a problem, but if part of a group, they will. If models are unable to analyse large groups of event logs, then they would be unsuitable for event log analysis because these events would go undetected, which poses a serious risk to system health and security. Our results in Figures \ref{fig:large_groups} and \ref{fig:all_events} demonstrate that the SLMs were able to analyse large groups of event logs, but all the LLMs (including Mistral 7b) were unable to analyse groups of logs that were larger than 7. The most plausible explanation for this is that the larger models are prioritising the instruction given to them rather than the previous response. This means that when they are instructed to either update the previous response with new information or keep it the same, they apply this to the instruction only, and the original prompt is disregarded after the first iteration. This is an important finding, as in a major category of our evaluation, the SLMs considerably outperform the LLMs in the domain of event log analysis.

\subsection{Model outputs}
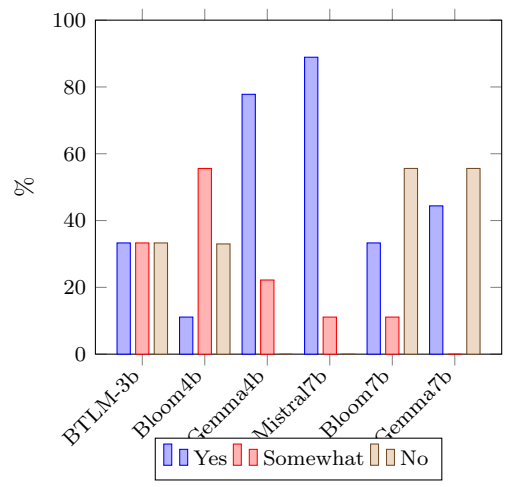
\begin{figure}[t]
\begin{tikzpicture}
\begin{axis}[
    ybar,
    bar width=5pt,
    height=6cm,
    ymin=0,
    ymax=100,
    ylabel={\%},
    ylabel style={font=\small},
    symbolic x coords={
        BTLM-3b,
        Bloom4b,
        Gemma4b,
        Mistral7b,
        Bloom7b,
        Gemma7b
    },
    xtick=data,
    xticklabel style={rotate=45, anchor=east, font=\footnotesize},
    yticklabel style={font=\footnotesize},
    legend style={
        font=\footnotesize,
        at={(0.5,-0.25)},
        anchor=north,
        legend columns=3
    },
    enlarge x limits=0.15
]

\addplot coordinates {
    (BTLM-3b,33.3)
    (Bloom4b,11.1)
    (Gemma4b,77.8)
    (Mistral7b,88.9)
    (Bloom7b,33.3)
    (Gemma7b,44.4)
};

\addplot coordinates {
    (BTLM-3b,33.3)
    (Bloom4b,55.6)
    (Gemma4b,22.2)
    (Mistral7b,11.1)
    (Bloom7b,11.1)
    (Gemma7b,0)
};

\addplot coordinates {
    (BTLM-3b,33.3)
    (Bloom4b,33)
    (Gemma4b,0)
    (Mistral7b,0)
    (Bloom7b,55.6)
    (Gemma7b,55.6)
};

\legend{Yes, Somewhat, No}

\end{axis}
\end{tikzpicture}
\caption{Distribution of responses to whether the outputs made sense.}
\label{fig:model_output}
\end{figure}
The evaluation form given to the experts contained a comprehensive set of questions to gather information about the output quality of the models. To begin, we will analyse information on whether the outputs made sense. This is an important metric as the problem-identified statements and solutions need to make sense for users and security personnel to act upon them. As seen in Figure~\ref{fig:model_output}, the experts agreed that Gemma 4b’s outputs made sense and were the best out of the SLMs; on the other hand, BTLM-3b and Bloom 4b largely made sense, but at times they did not. The experts unanimously agreed that Mistral 7b’s responses did make sense, but they were conflicted regarding Gemma 7b and Bloom 7b. Further analyse into the outputs and questioning some of the experts leads us to conclude that these two models often hallucinated severely and produced outputs that the experts questioned mentioned were the worst observed in this experiment. It is important to note that the pre-detection method rarely detected issues, and thus its impact was negligible. The experts mentioned that there were no differences between when an issue was pre-detected and when it was not.

\begin{figure}[t]
\begin{tikzpicture}
\begin{axis}[
    ybar,
    height=5.5cm,
    ymin=0,
    ymax=5,
    ylabel={Average Hallucination Rating},
    symbolic x coords={
        BTLM-3b,
        Gemma4b,
        Bloom4b,
        Mistral7b,
        Gemma7b,
        Bloom7b
    },
    xtick=data,
    xticklabel style={rotate=45, anchor=east, font=\footnotesize},
    yticklabel style={font=\footnotesize},
    bar width=8pt,
    enlarge x limits=0.15
]

\addplot coordinates {
    (BTLM-3b,3.33)
    (Gemma4b,2.33)
    (Bloom4b,4.11)
    (Mistral7b,1.78)
    (Gemma7b,3.56)
    (Bloom7b,3.89)
};

\end{axis}
\end{tikzpicture}
\caption{Average hallucination rating per model (lower indicates fewer hallucinations).}
\label{fig:hallucination_rating}
\end{figure}
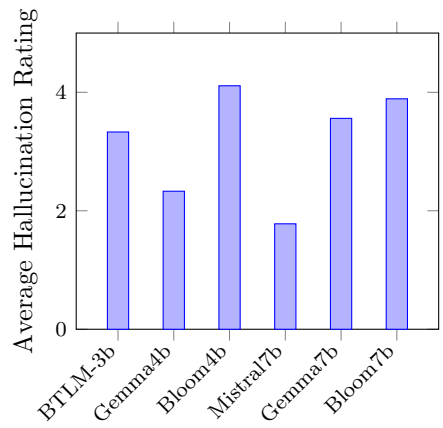

\begin{figure}[t]
\begin{tikzpicture}
\begin{axis}[
    ybar,
    height=6cm,
    ymin=0,
    ymax=100,
    ylabel={\%},
    symbolic x coords={
        BTLM-3b,
        Gemma4b,
        Bloom4b,
        Mistral7b,
        Gemma7b,
        Bloom7b
    },
    xtick=data,
    xticklabel style={rotate=45, anchor=east, font=\footnotesize},
    yticklabel style={font=\footnotesize},
    legend style={
        font=\footnotesize,
        at={(0.5,-0.25)},
        anchor=north,
        legend columns=3
    },
    bar width=5pt,
    enlarge x limits=0.15
]

\addplot coordinates {
    (BTLM-3b,66.7)
    (Gemma4b,88.9)
    (Bloom4b,44.4)
    (Mistral7b,88.9)
    (Gemma7b,55.6)
    (Bloom7b,55.6)
};

\addplot coordinates {
    (BTLM-3b,22.2)
    (Gemma4b,0)
    (Bloom4b,44.4)
    (Mistral7b,0)
    (Gemma7b,0)
    (Bloom7b,11.1)
};

\addplot coordinates {
    (BTLM-3b,11.1)
    (Gemma4b,11.1)
    (Bloom4b,11.1)
    (Mistral7b,11.1)
    (Gemma7b,44.4)
    (Bloom7b,33.3)
};

\legend{Yes, Somewhat, No}

\end{axis}
\end{tikzpicture}
\caption{Distribution of responses regarding understandability despite the presence of hallucinations.}
\label{fig:understanding}
\end{figure}
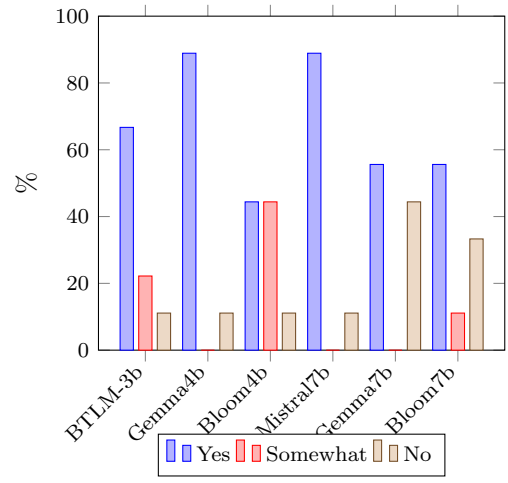

Adhering to the theme of hallucinations, we will discuss their frequency and impact. As demonstrated in figures \ref{fig:hallucination_rating} and \ref{fig:understanding}, the experts agreed that Gemma 4b and Mistral 7b had little to no hallucinations, and if they occurred, they could still understand the outputs. Bloom 4b and BTLM-3b often hallucinated, but the experts were still able to understand their outputs, as these hallucinations were a repetition of the input logs after the problem was identified, and solution statements were not within the statements themselves. On the other hand, as we mentioned previously, Gemma 7b and Bloom 7b severely hallucinated and often did not provide a relevant response to the logs provided.

\begin{figure}[t]
\begin{tikzpicture}
\begin{axis}[
    ybar,
    height=5.5cm,
    ymin=0,
    ymax=5,
    ylabel={Average Score},
    symbolic x coords={
        BTLM-3b,
        Gemma4b,
        Bloom4b,
        Mistral7b,
        Gemma7b,
        Bloom7b
    },
    xtick=data,
    xticklabel style={rotate=45, anchor=east, font=\footnotesize},
    yticklabel style={font=\footnotesize},
    bar width=8pt,
    enlarge x limits=0.15
]

\addplot coordinates {
    (BTLM-3b,2.89)
    (Gemma4b,4.22)
    (Bloom4b,2.22)
    (Mistral7b,4.11)
    (Gemma7b,2.00)
    (Bloom7b,1.78)
};

\end{axis}
\end{tikzpicture}
\caption{Distribution of responses regarding the models' ability to correctly identify issues.}
\label{fig:issue_identification}
\end{figure}
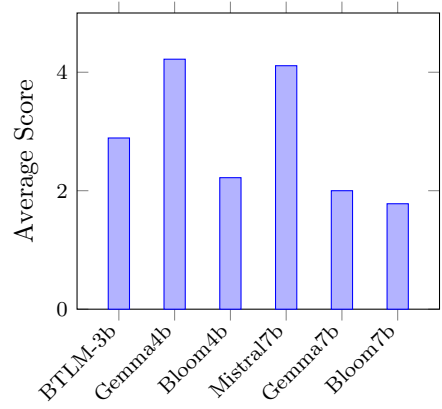

\begin{figure}[t]
\begin{tikzpicture}
\begin{axis}[
    ybar,
    height=6cm,
    ymin=0,
    ymax=100,
    ylabel={\%},
    symbolic x coords={
        BTLM-3b,
        Gemma4b,
        Bloom4b,
        Mistral7b,
        Gemma7b,
        Bloom7b
    },
    xtick=data,
    xticklabel style={rotate=45, anchor=east, font=\footnotesize},
    yticklabel style={font=\footnotesize},
    legend style={
        font=\footnotesize,
        at={(0.5,-0.25)},
        anchor=north,
        legend columns=3
    },
    bar width=5pt,
    enlarge x limits=0.15
]

\addplot coordinates {
    (BTLM-3b,55.6)
    (Gemma4b,66.7)
    (Bloom4b,44.4)
    (Mistral7b,77.8)
    (Gemma7b,44.4)
    (Bloom7b,11.1)
};

\addplot coordinates {
    (BTLM-3b,11.1)
    (Gemma4b,0)
    (Bloom4b,22.2)
    (Mistral7b,11.1)
    (Gemma7b,11.1)
    (Bloom7b,22.2)
};

\addplot coordinates {
    (BTLM-3b,33.3)
    (Gemma4b,33.3)
    (Bloom4b,33.3)
    (Mistral7b,11.1)
    (Gemma7b,44.4)
    (Bloom7b,66.7)
};

\legend{Yes, Somewhat, No}

\end{axis}
\end{tikzpicture}
\caption{Distribution of responses regarding whether the information provided was enough to resolve issues.}
\label{fig:info_sufficiency}
\end{figure}
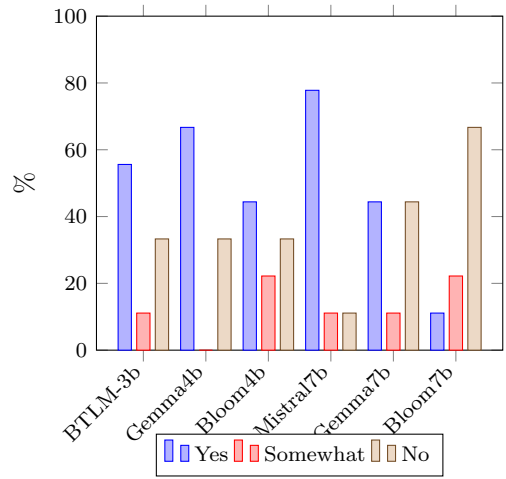

Moving on, the same trend is also observed concerning correctly identifying issues. As demonstrated in Figures \ref{fig:issue_identification} and \ref{fig:info_sufficiency}. For the SLMs, Gemma 4b was able to identify issues correctly and produce solutions that made sense, whereas Bloom 4b and BTLM 4b were slightly behind the model in this regard. For the LLMs Mistral 7b also identified issues correctly, and the solutions were in slightly more detail compared to Gemma 4b, whereas the other two LLMs were severely deficient in both, correctly identifying issues and providing detailed solutions.

As a result of this, we reach a surprising conclusion, which is that the SLMs (in general) produced better outputs compared to the LLMs for this task. This is because Bloom 7b and Gemma 7b consistently underperformed in all the categories that were analysed, and Gemma 4b was able to rival Mistral 7b in everything except for a slight decrease in output quality (whilst still being sufficient to resolve the issues and making sense). Another important conclusion is that we compared the SLM and LLM versions of Bloom and Gemma and found that the SLM versions produced better outputs. This could be because the larger models require a larger dataset to understand the task, whereas the smaller models have fewer parameters to change and can successfully learn with a small dataset. This is an important conclusion, as prior research into SLMs for event log analysis was limited to a single article where a model was used to aid an LLM, but our paper demonstrates that it is the better option, as our study affirms that a smaller dataset can be used successfully with SLMs for event log analysis tasks.  

\subsection{Best-performing model}
To conclude, we found that our objectives were met as the SLMs were shown to outperform the LLMs in all of our evaluation criteria, except for time taken, which could be addressed by splitting the tasks across multiple agents. Gemma 4b is the best-performing model, as it produced clear outputs that had little to no hallucinations and was able to analyse the large group of logs. There were only two disadvantages: the first was that the solutions it produced were slightly less detailed than Mistral, but were still sufficient and easy to understand. The second disadvantage was that it was not able to analyse the complete dataset, but this was likely since it was our last experiment and the GPU had been running for a considerable amount of time. In addition, this could be easily resolved by restarting the model and resuming the analysis. 

\subsection{Example pipeline}




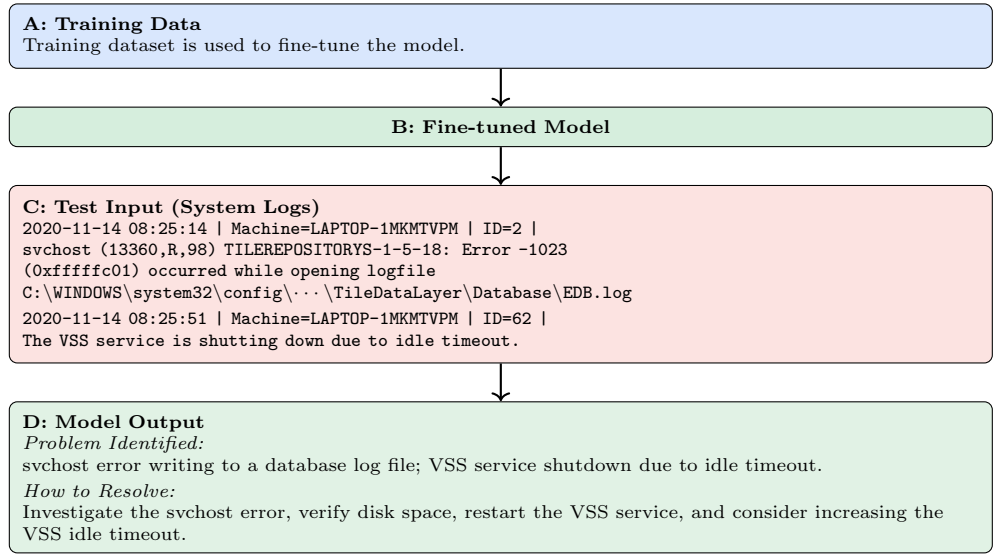
\begin{figure}[t]
\begin{tikzpicture}[
    node distance=0.5cm,
    box/.style={
        draw,
        rectangle,
        rounded corners=3pt,
        inner sep=5pt,
        align=left,
        text width=\columnwidth,
        font=\scriptsize
    },
    model/.style={
        draw,
        rectangle,
        rounded corners=3pt,
        inner sep=5pt,
        align=center,
        text width=\columnwidth,
        font=\scriptsize
    }
]

\definecolor{Acolor}{RGB}{66,133,244}
\definecolor{BCcolor}{RGB}{52,168,83}
\definecolor{Ccolor}{RGB}{234,67,53}

\node[box, fill=Acolor!20] (A) {
\textbf{A: Training Data}\\
Training dataset is used to fine‑tune the model.
};

\node[model, fill=BCcolor!20, below=of A] (B) {
\textbf{B: Fine‑tuned Model}
};

\node[box, fill=Ccolor!15, below=of B] (C) {
\textbf{C: Test Input (System Logs)}\\
\texttt{2020-11-14 08:25:14 | Machine=LAPTOP-1MKMTVPM | ID=2 |}\\
\texttt{svchost (13360,R,98) TILEREPOSITORYS-1-5-18: Error -1023}\\
\texttt{(0xfffffc01) occurred while opening logfile}\\
\texttt{C:\textbackslash WINDOWS\textbackslash system32\textbackslash config\textbackslash$\cdots$\textbackslash TileDataLayer\textbackslash Database\textbackslash EDB.log}\\[2pt]
\texttt{2020-11-14 08:25:51 | Machine=LAPTOP-1MKMTVPM | ID=62 |}\\
\texttt{The VSS service is shutting down due to idle timeout.}
};

\node[box, fill=BCcolor!15, below=of C] (D) {
\textbf{D: Model Output}\\
\textit{Problem Identified:}\\
svchost error writing to a database log file; VSS service shutdown due to idle timeout.\\[2pt]
\textit{How to Resolve:}\\
Investigate the svchost error, verify disk space, restart the VSS service, and consider increasing the VSS idle timeout.
};

\draw[->, thick] (A) -- (B);
\draw[->, thick] (B) -- (C);
\draw[->, thick] (C) -- (D);

\end{tikzpicture}
\caption{Diagram illustrating the fine‑tuned Gemma‑4B log analysis pipeline}
\label{fig:gemmapipeline}
\end{figure}

In Figure \ref{fig:gemmapipeline}, the pipeline for the fine-tuned Gemma 4b model is demonstrated. Firstly, the dataset was used to teach the model how to perform the task. Following fine-tuning, the testing dataset was sent to the model, and we highlight a problematic log that was present within the testing dataset. The model was able to identify that this was an issue and provided a detailed set of instructions to aid the user in resolving it. In addition, this set of instructions contains a description of the issue and how the suggestions resolve it. This will prevent users from making mistakes and could also help prevent a recurrence.

\section{Conclusion and future work}
In conclusion, we aimed to address the limitations of high resource consumption in fine-tuning and LLMs, and to expand research by producing a comprehensive tool that provides mitigative steps to address identified problems, rather than merely identifying them. To achieve these objectives, we generated a dataset which covered a wide range of activities, was akin to that which would be found in the real world, and included statements detailing the problems identified and providing clear steps to resolve them. In addition, we use the LoRA fine-tuning technique on a range of SLMs and LLMs to assess if SLMs could be used as an alternative, as they do not require high specification devices. We used experts to evaluate both the generated dataset and the fine-tuned models’ responses. The experts found that the generated dataset depicted real-world activities, correctly identified problems and provided detailed responses when problems were identified. For the second experiment, the experts made a surprising discovery, which was that the SLMs significantly outperformed the LLMs. This was clearly illustrated when comparing the SLM versions of Gemma and Bloom to their LLM counterparts. These points highlight that we successfully achieved our objective of expanding event log analysis research using language models by providing solutions to the problems identified, whilst reducing resource requirements and consumption. 

To further progress research, there are clear avenues that could be explored in future work. The first is to leverage the findings of this paper and produce multiple SLMs that are fine-tuned for specific types of event logs (for example, network logs or different operating systems such as Linux) in an agentic framework, which means that a singular event log analysis tool is required across an organisation. In addition, the limitations of the models’ responses could be addressed (such as the input logs being repeated in the output) by modifying the components of LoRA. Lastly, future research could further enhance the applicability of the proposed solution by developing a feature that allows the model to implement the provided solutions on behalf of and with the permission of users to mitigate the identified issues.

\bibliographystyle{plain}
\bibliography{bib}  

\end{document}